\documentclass[%
 reprint,
 superscriptaddress,
 amsmath,amssymb,
 aps,
floatfix,
]{revtex4-1}

\usepackage{graphicx}
\usepackage{dcolumn}
\usepackage{bm}

\newcommand{\degree}{\ensuremath{^\circ}}

\begin{document}

\preprint{APS/123-QED}

\title{Dynamic transition from $\alpha$-helices to $\beta$-sheets in polypeptide
superhelices}

\author{Kirill A. Minin}
\affiliation{Moscow Institute of Physics and Technology, Dolgoprudny, 141701, Russia}
\author{Artem Zhmurov}
\affiliation{Moscow Institute of Physics and Technology, Dolgoprudny, 141701, Russia}
\author{Kenneth A. Marx}   
\affiliation{Department of Chemistry, University of Massachusetts, Lowell, MA 01854, USA}   
\author{Prashant K. Purohit}
\affiliation{Department of Mechanical Engineering and Applied Mechanics, University of Pennsylvania, Philadelphia, PA 19104, USA}   
\author{Valeri Barsegov}
\affiliation{Department of Chemistry, University of Massachusetts, Lowell, MA 01854, USA}
\affiliation{Moscow Institute of Physics and Technology, Dolgoprudny, 141701, Russia}
   

\begin{abstract}
\noindent
We carried out dynamic force manipulations {\it in silico} on a variety of 
superhelical protein fragments from myosin, chemotaxis receptor, vimentin, fibrin, 
and phenylalanine zippers that vary in size and topology of their $\alpha$-helical packing. 
When stretched along the superhelical axis, all superhelices show elastic, plastic, and
inelastic elongation regimes, and undergo a dynamic transition from the $\alpha$-helices 
to the $\beta$-sheets, which marks the onset of plastic deformation. Using Abeyaratne-Knowles formulation of phase transitions, we developed a theory to model mechanical and kinetic 
properties of protein superhelices under mechanical non-equilibrium conditions and to map 
their energy landscapes. The theory was validated by comparing the simulated and theoretical 
force-strain spectra. Scaling laws for the elastic force and the force for $\alpha$-to-$\beta$ transition to plastic deformation can be used to rationally design new materials of required mechanical strength with desired balance between stiffness and plasticity.

\end{abstract}

\maketitle


In 1953 coiled-coils were proposed independently by Crick and Pauling 
as structures comprised of supercoiled $\alpha$-helical segments. Since 
then, coiled-coils have been recognized as ubiquitous, critically important, 
highly stable biomechanical structures, occuring either at the tissue 
level (hair, blood clots, etc.), in individual cellular structures 
(intracellular cytoskeleton, flagella, etc.), or as components of individual 
proteins. Many well studied proteins performing mechanical functions, 
utilize the superhelical coiled-coil architecture, including ones in the 
present study: muscle proteins (myosin), intermediate filaments (vimentin), 
blood clots and thrombi (fibrin), chemotaxis (chemotaxis receptors), 
cellular transport (kinesin), and bacterial adhesion (protein tetraprachion) 
\cite{LupasTBCS16}. Recently, the unique superhelical symmetry of coiled-coils 
has inspired the design of new materials \cite{QuinlanCOCB15}: short 
supercoils \cite{ArndtStructure02}, long and thick fibers \cite{PotekhinCB01}, 
nanotubes \cite{BurgessJACS15}, spherical cages \cite{FletcherScience13} 
and synthetic virions \cite{NobleJACS16}. In this study, we solve a longstandng 
problem, providing a theoretical basis for understanding different coiled-coils' 
stability and dynamic properties when undergoing the pulling force induced 
$\alpha$-helix to  $\beta$-sheet transition. We combined dynamic force manipulations 
{\it in silico} with theoretical modeling and found that all systems studied, 
with from two-to-five helices forming parallel and anti-parallel supercoil 
architectures, uniformly undergo three force induced extension regimes including 
a remarkable plastic phase transition from all $\alpha$-helices 
to all $\beta$-sheets. The quantitative agreement between the theory and 
simulations allows for a new approach to rationally design coiled-coils with 
specific mechanical properties into novel biomaterials applications.   

We used the atomic models (see Supplemental Material, 
SM) of myosin, vimentin, fibrin, bacterial chemotaxis receptor and 
phenylalanine zippers (PDB entries: 2FXO \cite{BlankenfeldtPNAS06}, 1GK4 
\cite{StrelkovEMBOJ02}, 3GHG \cite{KollmanBiochemistry09}, 1QU7 \cite{KimNature99}, 
2GUV and 2GUS \cite{LiuJMB06} respectively). (i) {\it Myosin II} contains 
a double-stranded parallel coiled-coil \cite{WarrickARCB87}. Upon muscle 
contraction, tension is transfered along the myosin tail \cite{WarrickARCB87} 
(experimental force data are available for myosin \cite{SchwaigerNatMater02,RootBPJ06}). 
(ii) {\it Vimentin}, in intermediate filaments in cells \cite{ErikssonJCI09,FletcherNature10}, helps determine their resistance to mechanical factors \cite{HerrmannNRMCB07}. 
Vimentin's structure contains an $\alpha$-helical rod domain, which can 
be divided into several double-helical parallel coiled-coil segments 
\cite{HerrmannNRMCB07}. (iii) {\it Fibrin} forms the fibrous network of a blood 
clot that stops bleeding \cite{Weisel05}. Triple-helical parallel coiled-coils 
create the unique visco-elastic properties of fibrin \cite{ZhmurovStructure11,ZhmurovJACS12}. 
(iv) {\it Bacterial chemotaxis receptor} is responsible for signal transduction 
across cell membranes \cite{KimNature99}. The cytoplasmic domain contains two 
double-stranded anti-parallel coiled-coils forming a four-stranded superhelix. 
(v) {\it Phenylalanine zippers}, artificial four-to-five stranded parallel 
supercoils, are promising biomaterials with tunable properties \cite{LiuJMB06,BurgessJACS15}.

{\it Force spectroscopy in silico:} We employed all-atom Molecular Dynamics (MD) 
simulations using the Solvent Accessible Surface Area (SASA) model with CHARMM19 
unified hydrogen force-field \cite{FerraraP04} implemented on a GPU (see SM) 
\cite{ZhmurovJACS12}. Protein models were constructed using the CHARMM program 
\cite{BrooksJCC09}. We used a lower damping coefficient $\gamma$$=$$0.15~ps^{-1}$ vs. $\gamma$$=$$50~ps^{-1}$ for ambient water at 300K for more efficient sampling 
of conformational space \cite{FalkovichPSSA10}. To mimic experimental conditions, 
we implemented the pulling plane with harmonically attached tagged residues at one 
end of the molecule and the resting plane with constrained residues at the other 
(Fig.~1a). The pulling plane was connected to a virtual cantilever moving with a 
velocity $v_f$$=$$10^4-10^6 \mu m/s$ and ramping up the force $f$$=$$r_{f}t$ with 
a loading rate $r_{f}$$=$$k_{s}v_{f}$$=$$10^{-3}$$-$$10^{-1}N/s$ ($k_{s}$$=$$100~pN/nm$ 
is the cantilever spring).


\begin{figure}[t]
\begin{center}
\includegraphics[width=1.0\linewidth]{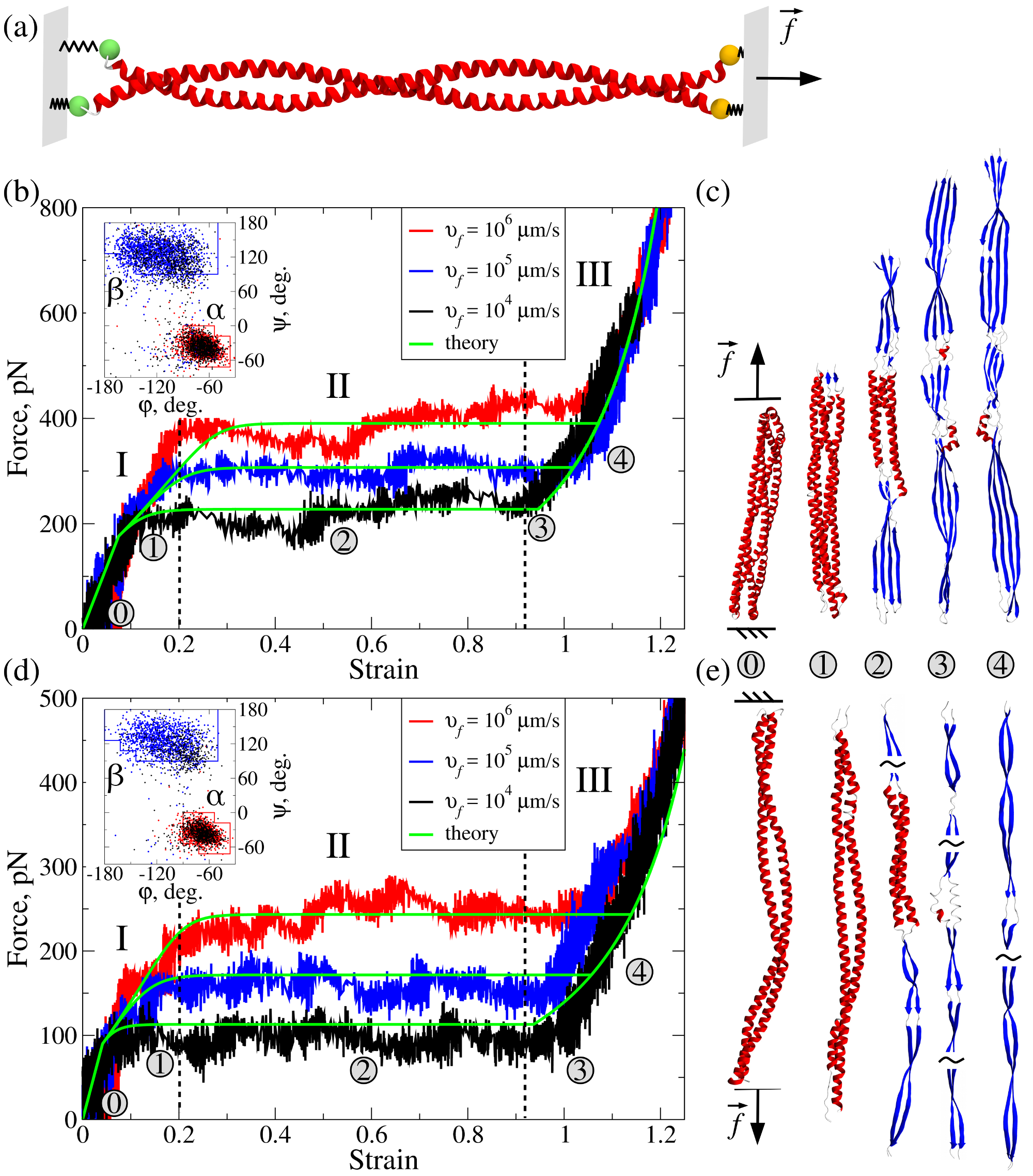}
\end{center}
\vspace{-0.6cm}
\caption{The $\alpha$-to-$\beta$ transition. 
{\it a:} pulling setup with pulling and resting planes. 
Transition is shown for the chemotaxis receptor 
({\it b}, {\it c}) and myosin ({\it d}, {\it e}). 
The $f$$\varepsilon$-curves ({\it b}, {\it d}) showing {\it elastic, plastic, and inelastic} regimes of superhelices' unfolding, overlaid with 
the continuum theoretical curves. {\it Insets} in {\it b} and {\it d} are 
Ramachandran plots of dihedral ($\phi$,$\psi$) angles in the two 
states. Snapshots ({\it c},{\it e}) from similarly numbered regions in $f$$\varepsilon$-curves show 
unfolding progress from initial $\alpha$-structure ($0$), to mixed 
$\alpha$$/$$\beta$-structures ($1$$-$$3$), to final $\beta$-structure ($4$).
Snapshots $3$$-$$4$ show exotic spirals of $\beta$-sheets.}
\label{fig:myosin-receptor-snap}
\vspace{-0.5cm}
\end{figure}


{\it The $\alpha$-to-${\beta}$ transition:} All superhelices undergo the 
force-driven transition from $\alpha$-helices to $\beta$-sheets regardless 
of the number of helices and parallel or anti-parallel architecture. 
In Fig.~1, we display the force-strain ($f\varepsilon$) curves (and structure 
snapshots) for bacterial receptor and myosin which are reminiscent of experimental 
force extension profiles \cite{SchwaigerNatMater02,RootBPJ06}. The plateau 
force for myosin $f^{*}$$\sim$$100~pN$ (Fig.~1d) is higher than the experimental 
value of $\sim$$40~pN$ \cite{RootBPJ06} due to $10^4$$-$$10^5$-fold faster 
pulling speeds used {\it in silico}. Results obtained for vimentin, fibrin and 
phenylalanine zippers are in Fig.~S1. The $f$$\varepsilon$-curves reflect the 
three regimes of dynamic mechanical response of the superhelices to an applied 
force: i) {\it elastic regime I} of coiled-coil elongation at low strain 
($f$$\sim$$\varepsilon$; $\varepsilon$$<$$0.2$), characterized by a linear growth 
of $f$ with $\varepsilon$; ii) {\it plastic transition regime II } of protein 
unfolding at intermediate strain ($f$$=$$\text{const}$; $0.2$$<$$\varepsilon$$<$$0.9$), 
where the $\alpha$-to-$\beta$ transition occurs; and iii) {\it inelastic regime III} 
of non-linear elongation of the $\beta$-structure at high strain 
($f$$\sim$$\varepsilon^2$; $\varepsilon$$>$$0.9$). In the $\alpha$-to-$\beta$ 
transition regime, the coiled-coils unwound and underwent a large $80$$-$$90$$\%$ 
elongation. The $\alpha$-to-$\beta$ transition nucleated at both ends of the 
molecule, and two phase boundaries propagated towards the center (snapshots 1--2, 
Fig.~1c,e). All of these features were observed irrespective of the number and 
architecture of $\alpha$-helices.

\begin{figure}[t]
\begin{center}
\includegraphics[width=0.90\linewidth]{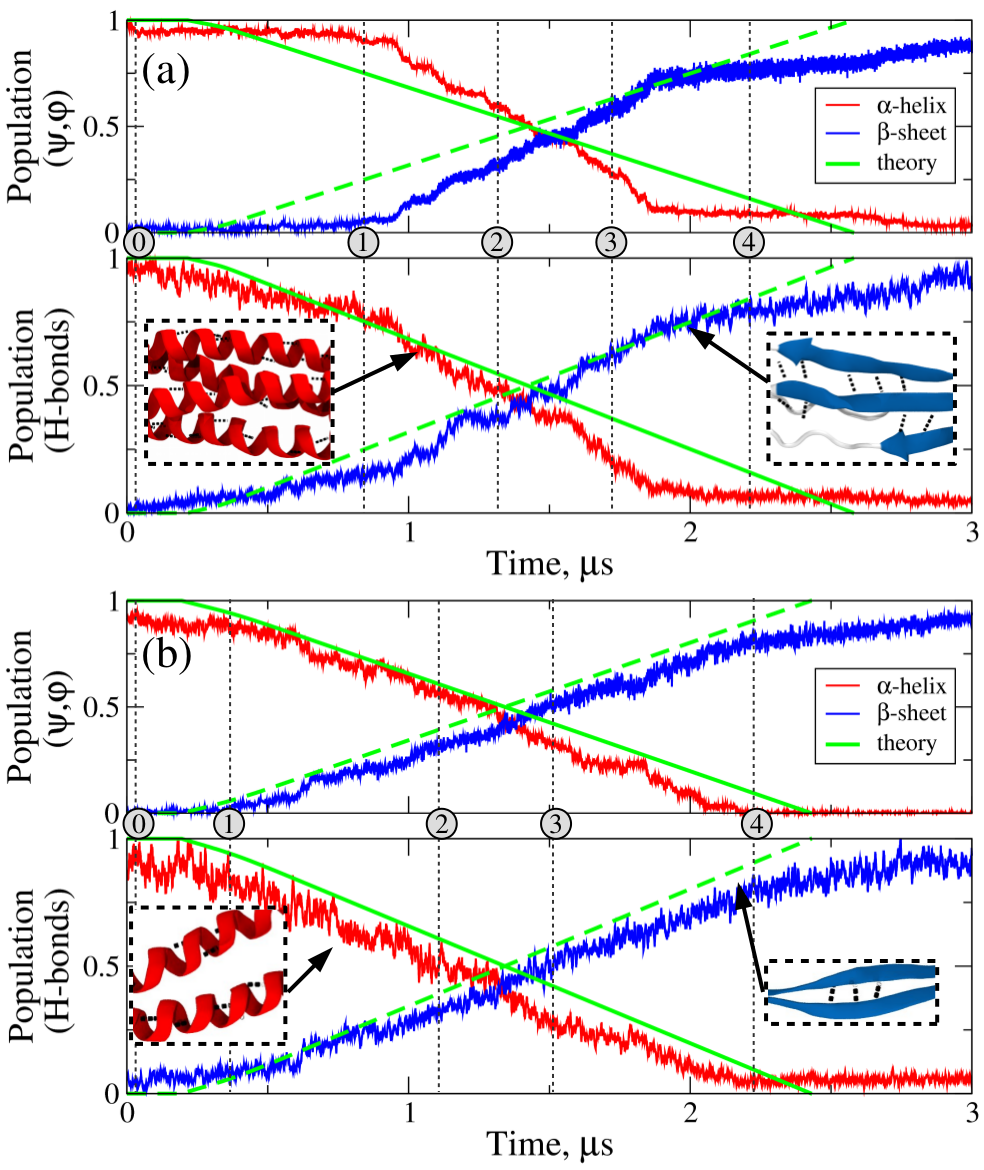}
\end{center}
\vspace{-0.6cm}
\caption{Dynamics of $p_{\alpha}$ and $p_{\beta}$ for chemotaxis 
receptor ({\it a}) and myosin tail ({\it b}), described using $\phi,\psi$-angles 
and H-bonds. Compared are curves from MD simulations 
with green curves from continuum theory. We used 
intervals $-80$$\degree$$<$$\phi$$<$$-48$$\degree$ and 
$-59$$\degree$$<$$\psi$$<$$-27$$\degree$ for the $\alpha$-phase, and $-150$$\degree$$<$$\phi$$<$$-90$$\degree$ and 
$90$$\degree$$<$$\psi$$<$$150$$\degree$ 
for the $\beta$-phase. \cite{SrinivasanProteins95}. For H-bonds, we used a 
$3$$\AA$ cut-off distance between hydrogen donor (D) and acceptor 
(A) atoms and a $20$$\degree$ cut-off for the D-H$\dots$A bond angle.}
\label{fig:props-hbonds-time}
\vspace{-0.5cm}
\end{figure}

Dihedral angles ($\phi$,$\psi$) are sensitive to changes in protein 
secondary structure \cite{SrinivasanProteins95}. This is reflected in 
migration of $\phi,\psi$-angles from the $\alpha$-region to the $\beta$-region 
in the Ramachandran plots ({\it insets} to Fig.~1b,d). Furthermore, in 
the $\alpha$-state H-bonds are all intramolecular while $\beta$-sheets 
form due to intermolecular H-bonds linking 
parallel or antiparallel $\beta$-strands. We profiled the probabilities 
of finding a system in the $\alpha$-state $p_{\alpha}$ and $\beta$-state 
$p_{\beta}$ using dihedral angles and H-bonds. We calculated 
the relative amount of intrachain bonds $p_{\alpha}$$=$$N_{intra}/N_H$ and 
interchain bonds $p_{\beta}$$=$$N_{inter}/N_H$ ($N_H$-total number of 
H-bonds), and the relative amounts of residues in the $\alpha$-region $p_{\alpha}$$=$$N_{\alpha}/N_{\phi\psi}$ and $\beta$-sheet region $p_{\beta}$$=$$N_{\beta}/N_{\phi\psi}$ ($N_{\phi\psi}$-total number of 
${\phi,\psi}$-angles). The profiles of $p_{\alpha}$ and $p_{\beta}$ displayed 
in Fig.~\ref{fig:props-hbonds-time} for bacterial receptor and myosin
show the following: (i) the $\alpha$-helical ($\beta$-strand) content 
decreases (increases) with time $t$ (and force $f$$=$$r_f$$t$); (ii) the 
transformation from $\alpha$-helices to $\beta$-sheets is a two-state 
transition; and (iii) this transition is accompanied by redistribution 
of $\phi,\psi$-angles ($\alpha$$\to$$\beta$; Fig.~1b,d {\it insets}) 
and reconfiguration of H-bonds (from intra- to interchain H-bonds; snapshots 
in Fig.~2). We obtained similar results for the vimentin, fibrin and 
Phe-zippers (Fig.~S2). Hence, the H-bonds and $\phi,\psi$-angles 
can be used as molecular signatures to characterize the $\alpha$-to-$\beta$ 
transition in proteins with superhelical symmetry.

{\it Two-state model:} Under the constant-force conditions (force-clamp), 
the $\alpha$-to-$\beta$ transition can be described using a two-state model. 
The sigmoidal extension-force phase diagram for myosin and fibrin coiled-coils 
in Fig.~S4 is divided into the $\alpha$-phase and $\beta$-phase \cite{ZhmurovJACS12}. 
The $\alpha$-state can be modeled as an entropic spring with energy 
$G$$=$$k_{\alpha}$$\Delta X_{\alpha}^2/2$, where $k_{\alpha}$ is the spring 
constant and $\Delta X_{\alpha}$$=$$f/k_{\alpha}$ is extension. The $\beta$-state 
can be described by a wormlike chain \cite{BarsegovBPJ06} with energy 
$G_{\beta}$$=$$(3k_BT/2l_{\beta})$$\int$$[\partial n(s)/\partial s]^2ds$, 
where $l_{\beta}$ is the persistence length, and $T$ is the temperature. 
A pulling force stretches the $\alpha$-helices by a fractional extension 
$y_{\alpha}(f)$$=$$\Delta X_{\alpha}(f)/L_{\alpha}$, where $L_{\alpha}$ 
is the maximal extension in the $\alpha$-state, and lowers the energy barrier 
$\Delta G(f)$ by increasing the transition probability 
$p_{\beta}(f)/p_{\alpha}(f)$$=$$\exp{[ -\Delta G(f)/k_BT]}$. The force 
induces elongation in the $\beta$-states, 
$y_{\beta}(f)$$=$$\Delta X_{\beta}(f)/L_{\beta}$, where $L_{\beta}$ is the 
maximal extension in the $\beta$-states. Here, 
$y_{\beta}(f)$$=$$1$$-$$\{\xi(q)^{1/3}$$+$$[4q/3-1]/\xi(q)^{1/3}\}^{-1}$, 
where $\xi(q)$$=$$2$$+$$\{4$$-$$[(4/3)q$$-$$1]^3\}^{1/2}$ and $q$$=$$fl_{\beta}/k_BT$. 
The total extension is
$\Delta X(f)$$=$$p_{\alpha}$$y_{\alpha}$$L_{\alpha}$$+$$p_{\beta}$$y_{\beta}$$L_{\beta}$ \cite{ZhmurovJACS12}.


{\it Continuum theory:} The $\alpha$-to-$\beta$ transition in the 
non-equilibrium regime of time-dependent force can be described using the 
Abeyaratne-Knowles formulation of phase transitions \cite{Abeyaratne06,RajEPL10}. 
The displacement of a material point at reference position $x$ at time $t$ 
is given by $u(x,t)$$=$$X(x,t)$$-$$x$. The end at $x$$=$$0$ is fixed, and 
$u(0,t)$$=$$0$ for all $t$. At the other end $x$$=$$L$, $u(L,t)$$=$$\Delta$$X(t)$, 
where $\Delta$$X(t)$$=$$v_ft$$>$$0$ ($<$$0$) when the sample is loaded (unloaded). 
The $\alpha$-phase is described by a stretch $\Gamma_{\alpha}(f)$$=$$X_{\alpha}(f)/L$, $f$$<$$f_{\alpha\beta}^*$, where $X_{\alpha}$ and $f_{\alpha\beta}^*$ are the 
end-to-end distance and critical force for the $\alpha$-phase. We set 
$L$$=$$L_{\alpha}$ throughout. At $f$$=$$f_{\alpha\beta}$, 
the $\beta$-phase nucleates and $f_{\alpha\beta}^*$$>$$f_{\alpha\beta}$$>$$f_0$, 
where $f_0$ is the Maxwell force at which the free energies per residue 
for the two phases become equal. The $\beta$-phase is described by a stretch 
$\Gamma_{\beta}(f)$$=$$X_{\beta}(f)/L$, $f_{\beta\alpha}^*$$<$$f$$<$$\infty$,
where $X_{\beta}(f)$ and $f_{\beta\alpha}^*$ are the end-to-end distance and 
critical force in the $\beta$-phase. A transformation strain is
$\gamma(f)$$=$$\Gamma_{\beta}(f)$$-$$\Gamma_{\alpha}(f)$, 
$f_{\alpha\beta}^*$$\geq$$f$$\geq$$f_{\beta\alpha}^*$. 

The equation of motion for the 1D continuum (assuming negligible drag and 
inertia forces) is $\partial f/\partial x$$=$$0$, so that $f(x,t)$ is constant 
for $0$$<$$x$$<$$L$. If $f_{\alpha\beta}^*$$<$$f$$<$$f_{\beta\alpha}^*$, a 
mixture of the $\alpha$- and $\beta$-phases is possible, and the total extension 
is given by
\begin{equation} \label{Eq2}
 X(L,t) = L + \Delta X(t) = L\left[\Gamma_{\alpha}(f(t))p_{\alpha}(t) + 
 \Gamma_{\beta}(f(t))p_{\beta}(t)\right]
\end{equation}
A kinetic relation expressed in terms of the force $f(t)$ describes the evolution 
of $p_{\alpha}$, i.e. 
\begin{equation}\label{Eq3}
\dot{p}_{\alpha}=\Phi(f(t)),
\end{equation}
where $\Phi(f)$ is a material property. By differentiating Eq.~(\ref{Eq2}) and 
eliminating $\dot{p}_{\alpha}$ using Eq.~(\ref{Eq3}), we obtain:
\begin{equation}\label{Eq5}
 \left[g(f) - \gamma'(f)(1+\Delta X/L)\right]\dot{f}
 + \gamma(f)v_{f}/L = \gamma^{2}(f)\Phi(f)
\end{equation}
where 
$g(f)$$=$$\Gamma_{\alpha}(f)\Gamma_{\beta}'(f)$$-$$\Gamma_{\alpha}'(f)\Gamma_{\beta}(f)$. 
From Eq.~(\ref{Eq5}) a force plateau ($\dot{f}$$=$$0$) forms at $f$$=$$f^*$ if either $\gamma(f^{*})=\Gamma_{\beta}(f^{*})-\Gamma_{\alpha}(f^{*})=0$, 
or
\begin{equation}\label{Eq6}
v_{f}/L = \gamma(f^{*})\Phi(f^{*}).
\end{equation} 
Because the height of force plateau depends on $r_f$ and, hence, is strain-rate ($v_{f}/L$) dependent, Eq.~(\ref{Eq6}) defines the force plateau height. 

{\it Application to $\alpha$-to-$\beta$ transition:} Calculating the $f$$\varepsilon$-curves requires a kinetic relation and a nucleation criterion. By treating the $\alpha$-phase as an entropic spring and the $\beta$-phase 
as a wormlike chain, we obtain $\Gamma_{\alpha}(f)=f/\kappa_{\alpha}$$ + 1$ and $\Gamma_{\alpha}'(f)$$=$$1/\kappa_{\alpha}$ for the $\alpha$-phase, where $\kappa_{\alpha}$$=$$k_{\alpha}/L_{\alpha}$, and 
$\Gamma_{\beta}(f)$$=$$(L_{\beta}/L_{\alpha})$$(1-\sqrt{k_BT/4l_{\beta}f})$ and 
$\Gamma_{\beta}'(f)$$=$$(L_{\beta}/4L_{\alpha})\sqrt{k_BT/l_\beta f^3}$ for the $\beta$-phase ($\Gamma_{\beta}(f)$ is obtained by inverting the wormlike chain formula for the force 
$f$ vs. extension $y_{\beta}$). The $\alpha$-to-$\beta$ transformation strain becomes:
\begin{equation}\label{Eq7}
\gamma(f)=(L_\beta/L_\alpha )\left( 1-\sqrt{k_BT/4l_\beta f}\right)-f/\kappa_\alpha - 1,
\end{equation}
and $\gamma'(f)$$=$$(L_\beta/4L_\alpha)\sqrt{k_BT/l_\beta f^3}$$-$$1/\kappa_\alpha$. 
To obtain the kinetic relation (Eq.~(\ref{Eq3})), we use the Arrhenius rates for 
forward and backward transitions $k_{\alpha\beta}$ and $k_{\beta\alpha}$ \cite{RajJMPS11},
\begin{equation}\label{Eq10}
\dot{p}_{\alpha} = \Phi(f) = A\left[e^{-\frac{\Delta G_{\alpha\beta}(f)}{k_BT}} 
- e^{-\frac{\Delta G_{\beta\alpha}(f)}{k_BT}}\right] 
\end{equation}
where $A$ is an attempt frequency, and 
\begin{equation}\label{Eq11}
\Delta G_{\alpha\beta}(f)=\epsilon_0 - fz_{\alpha\beta}, \quad 
\Delta G_{\beta\alpha}(f)=\epsilon_0 - \epsilon_{\alpha\beta} + fz_{\beta\alpha} 
\end{equation}
are the energy barriers. In Eq.~(\ref{Eq11}), $\epsilon_0$ and $\epsilon_{\alpha\beta}$ 
are the force-free energy barrier and energy difference (per residue length) between 
the $\alpha$- and $\beta$-states, respectively, and $z_{\alpha\beta}$ and $z_{\beta\alpha}$ 
are transition distances (see {\it the inset} to Fig.~3a). The detailed balance is given by:
\begin{equation}\label{Eq12}
\frac{p_{\beta}}{p_{\alpha}}=e^{-\frac{\Delta G_{\alpha\beta}(f)-\Delta G_{\beta\alpha}(f)}{k_BT}}=
e^{-\frac{\epsilon_{\alpha\beta}-f(z_{\alpha\beta}+z_{\beta\alpha})}{k_BT}}
\end{equation}


\begin{figure}[t!]
\begin{center}
\includegraphics[width=0.90\linewidth]{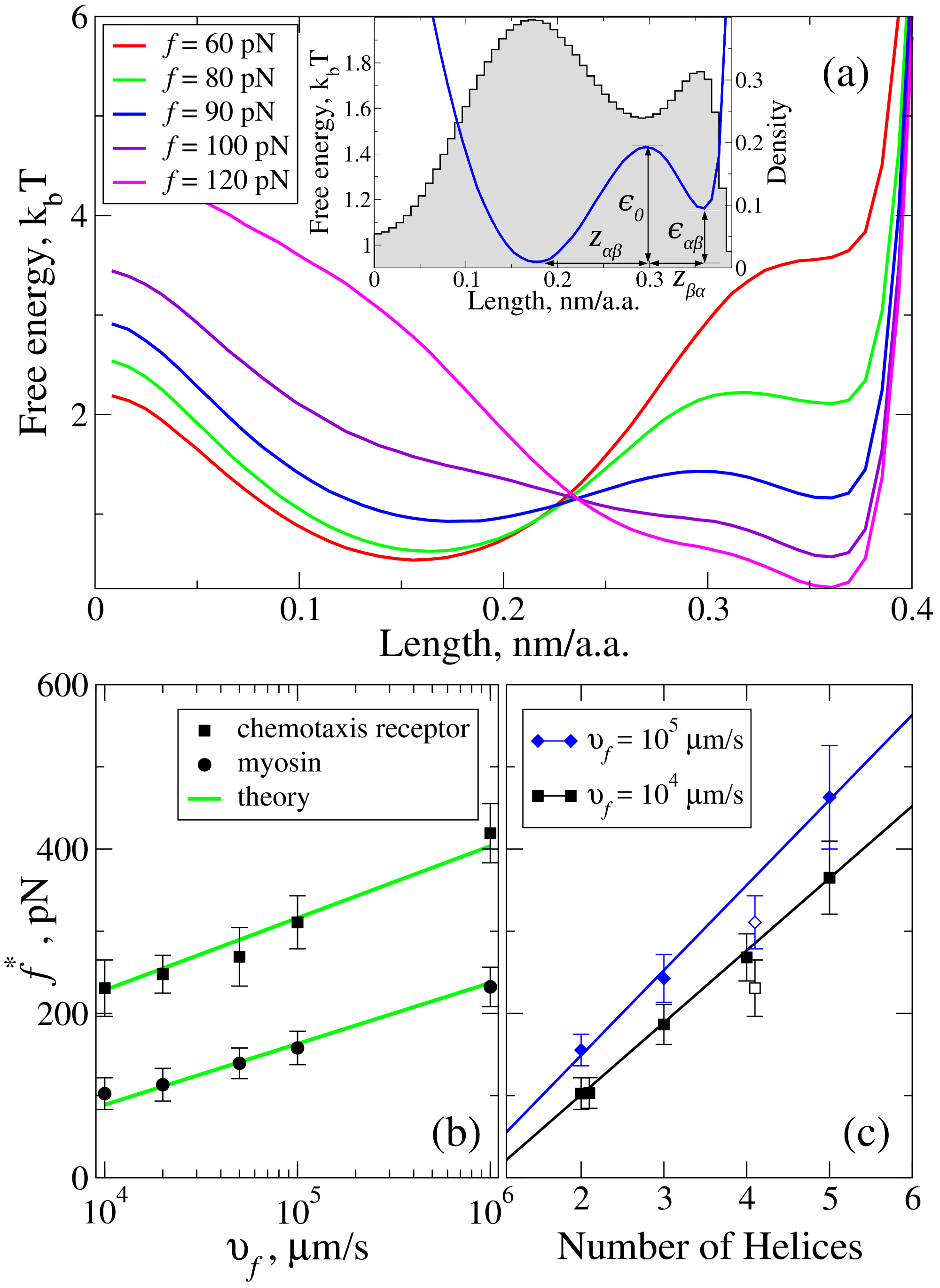}
\end{center}
\vspace{-0.6cm}
\caption{{\it a:} Free energy $G(x)$ for myosin superhelix at 
different forces. {\it Inset} shows a histogram of projected lengths $P(x)$
of myosin residues for $f$$=$$90~pN$ and energy landscape $G(x)$ with model parameters (Eqs. (\ref{Eq11})-(\ref{Eq13})). {\it b:} Plot of $f^*$ vs. $\ln v_f$ 
(data points) and theoretical curves of $f^*$ predicted using Eq.~(\ref{SEq5});
slopes and y-intercepts were calculated using parameters from Table~I. 
{\it c:} Plot of $f^*$ vs. $N_h$-number of $\alpha$-helices and a linear 
fit with $y$$=$$aN_h$ ($a$$=$$88$ and $103$ for $v_f$$=$$10^4$ and $10^5\mu$m/s).}
\label{fig:fstar-Nh}
\vspace{-0.5cm}
\end{figure}


\noindent
Setting $p_{\alpha}$$=$$p_{\beta}$ in Eq.~(\ref{Eq12}), we obtain the
Maxwell force:
\begin{equation}\label{Eq13}
f_0 = \epsilon_{\alpha\beta}/(z_{\alpha\beta} + z_{\beta\alpha})
\end{equation}

Initially, the continuum is in the $\alpha$-phase. Before the nucleation of 
the phase boundary, the force in the linear {\it elastic regime I} (Fig.~1b,d) 
is given by
\begin{equation}\label{Eq14}
f(t) = \kappa_{\alpha}v_{f}t/L_{\alpha} = 
\kappa_{\alpha}\Delta X_{\alpha}(t)/L_{\alpha}
\end{equation}
The nucleation criterion requires that when $f$$=$$f_{\alpha\beta}$, 
$\Delta X$$=$$\Delta X_{\alpha\beta}$$=$$f_{\alpha\beta}L/\kappa_{\alpha}$, a 
phase boundary nucleates at $x$$=$$0$ and $x$$=$$L$. The force is 
governed by Eq.~(\ref{Eq5}) with the initial condition $f$$=$$f_{\alpha\beta}$. 
The phase boundaries move through the continuum and convert all the material 
into the $\beta$-phase. This marks the onset of the {\it plastic transition 
regime II}, which corresponds to the force plateau in $f$$\varepsilon$-curves 
(Fig.~1b,d; see SM):
\begin{equation} \label{SEq5}
 f^{*}=\left(\frac{z_{\alpha\beta}}{k_{B}T}-\frac{L_{\alpha}}{\kappa_{\alpha}
 (L_{\beta}-L_{\alpha})}\right)^{-1}
\left[\ln\frac{v_{f}}{A\left(L_{\beta}-L_{\alpha}\right)}+\frac{\epsilon_{0}}{k_{B}T}\right]
\end{equation}
Using the structures generated for the mixed $\alpha$$+$$\beta$-phase, we 
run long $10~\mu s$ simulations for myosin and fibrin with pulling force 
gradually quenched to zero (with same $r_f$). The reverse $\beta$$\to$$\alpha$ 
transition was not observed. Hence, superhelices extended and $\beta$-sheets 
formed permanently, i.e. the $\alpha$$\to$$\beta$-transition marks the onset 
of plasticity in protein superhelices. Stretching of the $\beta$-phase continues 
in the {\it inelastic regime III} (Fig.~1b,d) and the force is given by the 
wormlike chain formula:
\begin{equation}\label{Eq15}
\begin{split}
f(t) &= (k_BT/4l_\beta)\left[1 - (L_{\alpha}/L_{\beta})
                       (1 + v_{f}t/L)\right]^{-2}\\
     &= (k_BT/4l_\beta)\left[1 - (L_{\alpha}/L_{\beta})
                       (1 + \Delta X_{\beta}(t)/L)\right]^{-2}
\end{split}
\end{equation} 

\begin{table*}[t]
\centering
\caption{Mechanical, thermodynamic and kinetic parameters for protein 
superhelices from myosin, chemotaxis receptor, vimentin, fibrin, and 
phenylalanine zippers differing in the number of $\alpha$-helices $N_h$: $L_{\alpha}$, $\kappa_{\alpha}$ and $z_{\alpha\beta}$ ($\alpha$-phase), $L_{\beta}$, $l_{\beta}$ and $z_{\beta\alpha}$ ($\beta$-phase), $f_0$, $\epsilon_{\alpha\beta}$, $\epsilon_{0}$ and $A$. Standard deviations (not shown) are $\le$$3\%$ of the average parameter values obtained 
directly from MD simulations or by using the energy landscapes (in parentheses), or 
by performing a fit of Eqs.~(\ref{Eq14})--(\ref{Eq15}) to the simulated 
$f$$\varepsilon$-curves in Figs.~1 and S1 (in squared brackets).}
\begin{tabular}{| c | c | c | c | c | c | c | c | c | c | c | c |}
\hline
Superhelical protein  & $N_h$ & $L_{\alpha}$, & $\kappa_{\alpha}$, & $l_{\beta}$, & $L_{\beta}$, & $z_{\alpha \beta}$, & $z_{\beta \alpha}$, & $f_0$,          & $\epsilon_{\alpha \beta}$, & $\epsilon_{0}$,  & $A$,            \\ 
fragments             &       & nm            & pN                 & nm           & nm           & nm                  & nm                  & pN              & $k_B T$                    & $k_B T$          & $\mu$s$^{-1}$   \\ \hline
Myosin                & $2$   & $18.1[18.1]$  & $2169[2170]$       & $[0.38]$     & $46[43]$     & $(0.13)[0.13]$      & $(0.07)[0.07]$      & $103(90)[90]$   & $(4.4)[4.4]$                    & $(3.3)[3.3]$          & $[0.5]$        \\
Chemotaxis receptor   & $4$   & $13.6[13.6]$  & $2367[2370]$       & $[0.33]$     & $34[32]$     & $(0.11)[0.11]$      & $(0.08)[0.08]$      & $232[176]$      & $(8.3)[8.3]$                    & $(5.3)[5.3]$          & $[0.2]$        \\
Vimentin              & $2$   & $11.5[11.5]$  & $1375[1375]$       & $[0.36]$     & $28[27]$     & $(0.14)[0.14]$      & $(0.06)[0.06]$      & $103[90]$       & $(4.3)[4.3]$                    & $(3.6)[3.6]$          & $[1.0]$        \\
Fibrin                & $3$   & $16.2[16.2]$  & $1135[1135]$       & $[0.35]$     & $41[38]$     & $(0.11)[0.11]$      & $(0.09)[0.09]$      & $187(160)[160]$ & $(7.8)[7.8]$                    & $(4.6)[4.6]$          & $[0.4]$        \\
4-strand. Phe-zipper  & $4$   & $6.2[6.2]$    & $2644[2645]$       & $[0.36]$     & $15[14]$     & $(0.11)[0.11]$      & $(0.08)[0.08]$      & $268[200]$      & $(9.3)[9.3]$                    & $(5.8)[5.8]$          & $[0.2]$       \\
Phe-zipper            & $5$   & $7.8[7.8]$    & $1793[1795]$       & $[0.35]$     & $20[19]$     & $(0.11)[0.11]$      & $(0.09)[0.09]$      & $383[350]$      & $(17)[17]$                   & $(9.4)[9.4]$          & $[0.1]$        \\ \hline
\end{tabular}
\label{tab:table1} 
\vspace{-0.3cm}
\end{table*}

\indent {\it Force spectra and energy landscape:} Parameters 
accessed directly is MD simulations are the end-to-end 
distances in the $\alpha$-phase $L_{\alpha}$; the slope of (elastic) 
portion of an $f$$\varepsilon$-curve $\kappa_\alpha$; the maximum extension 
in the $\beta$-phase $L_{\beta}$; and $f_0$$\approx$$f^*$ ($l_\beta$ is 
obtained using a wormlike chain fit to the inelastic part of a 
$f$$\varepsilon$-curve; see Fig.~\ref{fig:myosin-receptor-snap}). 
The other parameters, $z_{\alpha\beta}$, $z_{\beta\alpha}$, 
$\epsilon_{\alpha\beta}$, and $\epsilon_0$, are estimated by mapping 
the free-energy landscape. We employed a mean-field approach, 
$G(x)$$=$$-k_BT$$\ln P(x)$, to profile $G$ as a function of projection 
of the average residue length (unit length) along the superhelical 
axis $x$ (reaction coordinate). An example of $P(x)$ for 
myosin is in the {\it inset} to Fig.~3a (see Fig.~S3a for $P(x)$ 
and $G(x)$ for other superhelices). For $G(x)$ sampled
at $f$$\approx$$0$, $z_{\alpha\beta}$ and $z_{\beta\alpha}$ 
can be readily estimated. Since $z_{\alpha\beta}$ and $z_{\beta\alpha}$ 
barely change with $f$ (Fig.~3a), $\epsilon_{\alpha\beta}$ is obtained 
using the Maxwell force (Eq.~(\ref{Eq13})), and $\epsilon_0$ is obtained 
using either of Eqs.~(\ref{Eq11}). $A$ is estimated by fitting the profiles 
of $p_{\alpha}$ and $p_{\beta}$$=$$1$$-$$p_{\alpha}$ shown in Fig.~2 with 
Eq.~(\ref{Eq10}). All paramter values are presented in Table \ref{tab:table1}. 

We used the values of $L_{\alpha}$, $\kappa_{\alpha}$, $L_{\beta}$, 
$z_{\alpha\beta}$, $\epsilon_0$ and $A$ from Table \ref{tab:table1} as the 
input and Eqs. (\ref{Eq14})-(\ref{Eq15}) to fit the simulated 
$f$$\varepsilon$-curves for all the protein superhelices at all 
pulling speeds (in Eqs. (\ref{Eq14}) and (\ref{Eq15}), 
$\varepsilon(t)$$=$$\Delta$$X(t)/L$ is the time-dependent
strain). This enabled us to refine the parameter values and estimate 
$l_{\beta}$ (shown in squared brackets in Table \ref{tab:table1}). 
The results of fitting for chemotaxis receptor and myosin presented in 
Fig.~1b,d and in Fig.~S1 for the other superhelices show excellent agreement 
between simulated and theoretical $f$$\varepsilon$-curves, which 
validates our theory. The refined final values of model parameters from 
the fit do not differ much from the initial input, which points to the 
internal consistency between the results of theory and simulations. We 
used these parameters and Eq.~(\ref{Eq10}) to reconstruct the profiles 
$p_{\alpha}$ and $p_{\beta}$. The Fig.~2 results show that, although 
H-bond redistribution and dihedral angles' migration capture the 
$\alpha$-to-$\beta$ transition, the H-bond based estimation of $p_{\alpha}$ 
and $p_{\beta}$ results in a better agreement with the simulations.

\indent {\it Scaling laws:} The theory predicts that the stiffness of 
superhelices $\kappa_{\alpha}/L_{\alpha}$ is inversely proportional to their 
length $L_{\alpha}$ (Eqs. (\ref{Eq14})), so shorter (longer) superhelices
are stiffer (less stiff). It also predicts a logarithmic scaling of the 
critical force $f^*$ with the pulling speed $v_f$ (loading 
rate $r_f$$=$$k_s$$v_f$; Eq.~(\ref{SEq5})). We tested this prediction by 
performing a fit of the $f^*$ vs. $\ln v_f$ profiles for superhelices from 
chemotaxis receptor and myosin with Eq.~(\ref{SEq5}) borrowing parameter values 
from Table \ref{tab:table1}. The Fig.~3b results show excellent agreement 
between the predicted and simulated values of $f^*$ extracted from the 
$f$$\varepsilon$-spectra (Figs.~1 and S1), which further validates our theory. 
Also, $f^*$$\sim$$N_h$ --- the number of $\alpha$-helices forming a superhelix 
($N_h$$=$$2$$-$$5$; Table~\ref{tab:table1}). The profiles of $f^*$ vs. 
$N_h$ in Fig.~3c show a roughly additive contribution to the mechanical 
strength from $\alpha$-helices, which weakly cooperate to sustain the stress. 
For $v_f$$=$$10^5\mu$m/s, $f^*$ increases from $103~pN$ for double-helical 
myosin tail and vimentin ($\sim52~pN$ per helix) to $187~pN$ for the 
triple-helical fibrin coiled-coil ($\sim$$62~pN$ per helix), to $268~pN$ 
for four-stranded Phe-zipper ($\sim$$67~pN$ per helix), and to $383~pN$ 
for Phe-zipper with five $\alpha$-helices ($\sim$$77~pN$ per helix). 
A lower $f^*$$=$$232~pN$ for the four-stranded chemotaxis receptor 
($\sim$$58~pN$ per helix) is due to the antiparallel arrangement implying
that parallel arrangement of $\alpha$-helices provides higher 
mechanical strength.

The developed theory can be used to accurately describe dynamic transitions 
in wild-type and synthetic superhelical polypeptides under mechanical
non-equilibrium conditions and model their force-strain spectra from dynamic 
force experiments. The theory can be used to probe mechanical and kinetic 
characteristics of any coiled-coil superhelical polypeptide and to map out their 
free-energy landscapes. The slope and $y$-intercept of the line of critical 
force $f^*$ vs. $\ln v_f$ (Fig. 3b) can be used to estimate the critical 
distance $z_{\alpha\beta}$ and the force-free barrier height $\epsilon_0$ 
(SM). Scaling laws for the elastic force $f$ vs.length $L$, and $f^*$ vs. 
loading rate $r_f$, and the dependence of $f^*$ on $L$ and number of helices 
$N_h$ provide a powerful new method to rationally design novel synthetic 
biomaterials with the required mechanical strength and balance between stiffness 
and plasticity.


\noindent
{\bf Acknowledgments:} This work was supported by NSF (grant DMR1505662 to 
VB and PKP), American Heart Association (grant-in-aid 13GRNT16960013 to VB), 
and Russian Foundation for Basic Research (grants 15-37-21027, 15-01-06721 
to AZ).


\bibliography{coiledcoils}
\end{document}


\preprint{APS/123-QED}

\title{Dynamic transition from $\alpha$-helices to $\beta$-sheets in polypeptide 
superhelices}

\author{Kirill A. Minin}
\affiliation{Moscow Institute of Physics and Technology, Dolgoprudny, 141701, Russia}
\author{Artem Zhmurov}
\affiliation{Moscow Institute of Physics and Technology, Dolgoprudny, 141701, Russia}
\author{Kenneth A. Marx}   
\affiliation{Department of Chemistry, University of Massachusetts, Lowell, MA 01854, USA}   
\author{Prashant K. Purohit}
\affiliation{Department of Mechanical Engineering and Applied Mechanics, University of Pennsylvania, Philadelphia, PA 19104, USA}   
\author{Valeri Barsegov}
\affiliation{Department of Chemistry, University of Massachusetts, Lowell, MA 01854, USA}
\affiliation{Moscow Institute of Physics and Technology, Dolgoprudny, 141701, Russia}
   

\begin{abstract}
\noindent
\begin{center}
\Large Supplemental Material
\end{center}
\end{abstract}

\maketitle


\section{Model systems}

In this study, we have selected several coiled-coils fragments from various proteins (Table~\ref{tab:systems}). The idea was to cover all known variants of the coiled coil geometries, including both parallel and anti-parallel arrangements of the $\alpha$-helices in supercoils and differing numbers of $\alpha$-helices 
$N_h$$=$$2$$-$$5$ (see Fig.~\ref{fig:s5} and Table I in the Main Text). We deliberately selected proteins whose function has a mechanical nature. 
Also, we have added a couple of synthetic coiled-coils since these are 
very promising systems in terms of material design applications. All 
the superhelical protein fragments used in this work are summarized in Table~\ref{tab:systems}. Structure snapshots for all superhelices are
presented in Figs.~S1 and S3.

\begin{table*}[t]
\centering
\caption{Structure and topology of superhelical proteins: the PDB entries 
(four-character codes) of the atomic structural models available from the 
Protein Data Bank; $N_h$ is the number of $\alpha$-helices formulating the 
superhelical structure; the vertical arrows show the topology and mutual 
orientation of the $\alpha$-helices in space; stretches of amino acids 
provide information about the identity of $\alpha$-helical residues formulating
the superhelical architecture (taken from PDB files).}
\vspace{1cm}
\begin{tabular}{| l | c | c | c | c | c | c |}
\hline
System                     & PDB  & $N_h$ & Topology                                   & Residues                                          &                      \\  \hline
Myosin tail                & 2FXO & $2$   & $\uparrow\uparrow$                         & 835--963, 831--961                                & Homo-dimer           \\
Vimentin                   & 1GK4 & $2$   & $\uparrow\uparrow$                         & 328--406, 328--406                                & Homo-dimer           \\
Fibrinogen coiled-coils    & 3GHG & $3$   & $\uparrow\uparrow\uparrow$                 & $\alpha$49--161, $\beta$80--193, $\gamma$23--135  & Hetero-trimer        \\
Four-stranded Phe-zipper   & 2GUS & $4$   & $\uparrow\uparrow\uparrow\uparrow$         & (13-54)$\times$4                                  & Homo-tetramer        \\
Bacterial receptor         & 1QU7 & $4$   & $\uparrow\downarrow\uparrow\downarrow$     & 294--489, 300-479                                 & Homo-dimer           \\
Phenylalanine zipper       & 2GUV & $5$   & $\uparrow\uparrow\uparrow\uparrow\uparrow$ & (1-56)$\times$5                                   & Homo-pentamer        \\ \hline
\end{tabular}
\label{tab:systems} 
\end{table*}

{\bf Myosin} is a muscle protein containing long double-stranded coiled-coil in its tail fragment \cite{WarrickARCB87}. Upon muscle contraction, the head of the myosin moves along the actin filament and the resulting mechanical tension is transferred along the tail \cite{WarrickARCB87}. In this study, we used the structure of the part of the human myosin II tail (PDB structure 2FXO \cite{BlankenfeldtPNAS06}). This structure contains two chains with residues Gly835-Lys963 and Ser831-Leu961, respectively, that form a double-stranded homodimeric parallel coiled-coil. There is also a good amount of experimental data available for myosin coil, including the
force-extension curves from single-molecule AFM experiments \cite{SchwaigerNatMater02,RootBPJ06}. Myosin coiled-coils can be classified as 
leucine zippers, with around 50\% of amino-acids on the positions {\bf a} and {\bf d} being leucine residues (Fig.~\ref{fig:s5} and Section~\ref{sec:cc-arch}). The second most popular amino 
acid at these positions is methionine (around 25\%).

{\bf Vimentin} is an elementary building block of intermediate filaments from connective tissue. It is also found in other types of mesodermal tissue \cite{ErikssonJCI09}. Intermediate filaments, along with microtubules and actin, are involved in the construction of the cytoskeleton \cite{FletcherNature10}. Vimentin plays a significant role in fastening organelles and maintaining their positions in the cytoplasm. While most of the intermediate filaments retain their structure, vimentin filaments are dynamic, which is important when the cell changes its shape. Vimentin provides strength to the cells, thereby helping to determine their resistance to mechanical stress and to maintain their integrity \cite{HerrmannNRMCB07}. The mechanical function of the vimentin makes it a natural system for our study. Similarly to many other intermediate filaments, the structure of vimentin can be divided into three domains --- non-$\alpha$-helical head and tail domains and an $\alpha$-helical rod domain. While head and tail domains are responsible for the lateral aggregation, the mechanical footprint of vimentin is solely determined by the $\alpha$-helical coiled-coil of the rod domain \cite{HerrmannNRMCB07}. This domain can be divided into several coiled-coils segments. The atomistic structure of some of these segments was determined by X-Ray crystallography \cite{StrelkovEMBOJ02}. In this study, we used the structure of the 2B segment from human vimentin, namely residues Cys328 to Gly406 in each of two $\alpha$-helical strands (PDB structure 1GK4 \cite{StrelkovEMBOJ02}). Together, these helices form a parallel two-stranded coiled-coil, mostly stabilized by leucine residues on {\bf a} and {\bf d} positions (see Fig.~\ref{fig:s5} and Section~\ref{sec:cc-arch}).

{\bf Fibrinogen} is an abundant blood protein. Upon conversion to fibrin it forms the fibrous network called fibrin gel \cite{FerryPNAS52,Weisel05}. This network acts as a scaffold for blood clot formation, capturing platelets and other blood cells to form a plug that stops bleeding. It is hypothesized that both globular parts and coiled-coils are responsible for the unique mechanical properties of fibrin \cite{ZhmurovStructure11,ZhmurovJACS12,WeiselScience08,BrownBPJ07}. In this study, we used the structure of the entire human fibrinogen molecule (PDB structure 3GHG \cite{KollmanBiochemistry09}) and isolated the coiled-coils region, which contain residues $\alpha$Cys49-Cys161, $\beta$Cys80-Cys193 and $\gamma$Cys23-Cys135. The fibrinogen coiled-coil is a parallel heterotrimeric left-handed supercoil (Fig.~\ref{fig:s5} and Section~\ref{sec:cc-arch}). An interesting feature of this particular structure is the presence 
of the plasmin binding site in the center --- a slight distortion in the $\alpha$-helical structure which also forms a flexible hinge \cite{KohlerPLoS15}.

{\bf The bacterial chemotaxis receptors} are transmembrane receptors with a simple signaling pathway responsible for signal recognition and transduction across membranes \cite{WadhamsNRMCB04}. In contrast to many mammalian receptors, whose signaling mechanism involves forming protein oligomers upon ligand binding, the chemotaxis receptors are dimeric even in the absence of their ligands, and their signaling does not depend on an equilibrium between monomeric and dimeric states. Bacterial chemotaxis receptors are composed of a ligand-binding domain, a transmembrane domain consisting of two helices TM1 and TM2, and a cytoplasmic domain. All known bacterial chemotaxis receptors have a highly conserved cytoplasmic domain, which unites signals from different ligand domains into a single signaling pathway \cite{KimNature99}. The cytoplasmic domain contains two identical chains that make a ``U-turn'' at the end that is furthest from the membrane. These two chains form a two double-stranded anti-parallel coiled coil that are wrapped around one another to form a four-stranded supercoil (Fig.~\ref{fig:s5} and Section~\ref{sec:cc-arch}). In this work we use the part of this structure that is bordered by the residues Arg294-Ala489 for the first and Ala300-Glu479 for the second chain \cite{KimNature99}.
   
{\bf Phenylalanine zippers} are artificial proteins that resemble five- or four stranded parallel left-handed coiled coils (PDB structures 2GUV and 2GUS \cite{LiuJMB06}). As the name suggests, the hydrophobic {\bf a} and {\bf d} positions are occupied by phenylalanine residues (Fig.~\ref{fig:s5} and Section~\ref{sec:cc-arch}). In the coiled-coils 
these residues form a hydrophobic core, interlocking in a ``knobs-into-holes'' pattern. Interestingly, changing of only one Phe27 to a Met27 residue favors the formation of tetramer (PDB structure 2GUS) instead of the pentamer (PDB structure 2GUV). Although these coiled-coils are synthetic, the zippers they form are very promising building blocks for design of new materials, including those with tunable mechanical properties \cite{KohnTBT98,PotekhinCB01,BurgessJACS15}. In this work, we used the entire homopentameric (homotetrameric) structures, which involve residues Ser1-Arg56 in 
each of the five (four) chains.


\section{MD simulations}

{\bf Pulling simulations:} In our computer-based modeling, we utilized our 
in-house code for the all-atom MD simulations in implicit water \cite{ZhmurovJACS12} fully implemented on a Graphics Processing Units (GPUs). We used the SASA model of implicit solvation in conjunction with the CHARMM19 unified hydrogen force field \cite{FerraraP04}. To speedup simulations, all the computational subroutines 
were fully implemented on a GPU. This enabled us to attain long $1$$-$$10$$\mu s$-timescales in reasonable wall-clock time. All the model systems were built 
using the CHARMM program \cite{BrooksJCC09}. The systems were energy-minimized 
using the steepest descent algorithm and gradually heated up to $T$$=$$300 K$ 
in a span of $300~ps$. We used lower damping with a smaller damping coefficient $\gamma$$=$$0.15$~$ps^{-1}$ compared to $\gamma$$=$$50$~$ps^{-1}$ for ambient 
water at $T$$=$$300~K$. In effect, lower damping (viscosity) allows for faster 
and more effivient sampling of the conformational phase space \cite{FalkovichPSSA10,LiangBPJ06,WestJCP06}. A similar approach was used, e.g., 
in Ref.~\cite{FalkovichPSSA10}.

Protein superhelices are expected to unwind in the course of mechanical unfolding \cite{ZhmurovJACS12}. The process of unwinding is accompanied by rotational 
motion of certain atoms in the plane perpendicular to the pulling axis 
(superhelical symmetry axis). In the standard Steered Molecular Dynamics (SMD) approach, there is a designated set of fixed atoms and a set of atoms that are pulled. The pulling force is determined by the Hooks Law with a virtual spring attached to the pulled atoms. The other end of the spring moves with a constant velocity mimicking the cantilever motion in the AFM experiment. There are problems with this approach when it is applied to the coiled-coil proteins. First, the fixed atoms cannot move relative to one another, which restricts their rotational motion \cite{RootBPJ06}. Second, the force is applied to all $\alpha$-helical chains independently, which might cause some chains' sliding past one another. 

{\bf Force protocol:} To overcome this problem and to ensure an equal 
distribution of the force load, while also allowing the rotational motion 
of atoms around the pulling axis, we have implemented the following setup. 
We introduced two planes that are perpendicular to the pulling direction. 
One of the planes was fixed, representing a constrained end of a protein system. 
The other plane was pulled using the external force (see Fig.~1a in the Main 
Text). The selected constrained and tagged atoms for each superhelical system 
were attached to the corresponding plane via harmonic springs. The forces 
on the springs were acting only in the longitudinal direction perpendicular 
to the planes, thus allowing atoms to move freely in the transverse directions. 
The pulled plane was connected to a virtual cantilever bead that moved with the constant velocity $v_f$, ramping up the force $f$$=$$r_{f}t$ with the force-loading rate $r_{f}$$=$$k_s$$v_{f}$. 

We solved numerically the overdamped Langevin equation of motion for the pulled plane with the timestep $\tau_{pl}$$=$$100$$\times$$\tau$$=$$100~fs$, where $\tau$$=$$1~fs$ 
is the timestep used in the numerical integration of the Langevin equations of motion 
for the atoms. The cantilever spring constant $k_{s}$ was set to $100$~$pN/nm$ 
in all the simulations described in this work. For each model system, we performed 
5 simulation runs for each of the pulling speeds $v_f$$=$$10^4$, $10^5$ and $10^6\mu m/s$ (force-ramp simulations). We also performed the constant-force pulling experiments (force clamp simulations) using the following force values for the myosin tail: $f$$=$$20$, $40$, $60$, $70$, $80$, $90$, $100$, $120$, $130$, $140$, $160$, $180$ and $200$ pN, and for fibrin: $f$$=$$40$, $60$, $80$, $100$, $120$, $140$, $160$, $180$, $200$, $220$, $240$, $260$ and $300$ pN.


\section{Derivation of Equation (11)}

The dynamic force-ramp experiments {\it in silico} produced force plateaus that 
are mostly flat, albeit a bacterial receptor produces a slightly downward 
sloping plateau. When we plug in parameters $\kappa_{\alpha}$, $L_{\beta}/L_{\alpha}$, etc., from Table~1 in the Main Text into the transformation strain,$\gamma(f^{*})$$=$$\Gamma_{\beta}(f^{*})$$-$$\Gamma_{\alpha}(f^{*})$$=$$0$, 
we obtain much lower values of the plateau forces than those obtained in the all-atom MD simulations. Hence, the plateau forces $f^{*}$ conform to Eq.~(4) in the Main Text, and also imply that $f^{*}$ is a solution of the following algebraic 
equation:
\begin{equation}\label{SEq1}
\begin{split}
 A&\left[\exp(-\frac{\epsilon_{0} - f^{*}z_{\alpha\beta}}{k_{B}T})
           - \exp(-\frac{\epsilon_{0} - \epsilon_{\alpha\beta} 
           + f^{*}z_{\beta\alpha}}{k_{B}T})\right]\\
           \times &\left[\frac{L_{\beta}}{L_{\alpha}}\left(1 
                   - \sqrt{\frac{k_{B}T}{4l_{\beta}f^{*}}}\right)
                   - \frac{f^{*}}{\kappa_{\alpha}} - 1\right] = \frac{v_{f}}{L}.
\end{split}
\end{equation}
The meaning of all model parameters entering Eq.~(\ref{SEq1}) is described in the
Main Text (see also Table I in the Main Text). For large values of $f^{*}$, the second exponential in the first square brackets 
in Eq.~(\ref{SEq1}) becomes negligible compared to the first exponential, 
and so the second exponential can be neglected. Also, 
$\sqrt{\frac{k_{B}T}{4l_{\beta}f^{*}}}$ tends to zero as $f^{*}$ increases, 
and so we can neglect this term in comparison with the term 
$f^{*}/\kappa_{\alpha}$ in the second square bracket. Hence, we obtain from
Eq.~(\ref{SEq1}) the following approximate equation for $f^{*}$: 
\begin{equation} \label{SEq2}
 \exp{[-\frac{\epsilon_{0} - f^{*}z_{\alpha\beta}}{k_{B}T}]}
           \left[\frac{L_{\beta}}{L_{\alpha}} - \frac{f^{*}}{\kappa_{\alpha}} - 1\right]
 = \frac{v_{f}}{AL}
\end{equation}
By taking the logarithm on both sides of Eq.~(\ref{SEq2}) above we obtain the following equation:
\begin{equation}\label{SEq3}
 -\frac{\epsilon_{0} - f^{*}z_{\alpha\beta}}{k_{B}T}
           + \ln\left(\frac{L_{\beta}}{L_{\alpha}} - \frac{f^{*}}{\kappa_{\alpha}} - 1\right) 
 = \ln\frac{v_{f}}{AL}.
\end{equation}
Assuming that $\frac{f^{*}L_{\alpha}}{\kappa_{\alpha}(L_{\beta}-L_{\alpha})} < 1$, based on the results of MD simulations, we can expand the logarithmic term on the left-hand-side of the above Eq.~(\ref{SEq3}) in Taylor series. By retaining the linear terms in the Taylor expansion, we obtain:
\begin{equation}\label{SEq4}
 f^{*}\left(\frac{z_{\alpha\beta}}{k_BT}-\frac{L_{\alpha}}{\kappa_{\alpha}(L_{\beta}-L_{\alpha})}\right)=\ln\frac{v_{f}}{A\left(L_{\beta}-L_{\alpha}\right)}+\frac{\epsilon_{0}}{k_{B}T}, 
\end{equation}
Solving Eq.~(\ref{SEq4}) for the plateau force $f^*$, we obtain the final expression 
for the calculation of $f^*$, which reads:
\begin{equation} \label{SEq5}
 f^{*} = \left(\frac{z_{\alpha\beta}}{k_{B}T} - \frac{L_{\alpha}}{\kappa_{\alpha}
 (L_{\beta}-L_{\alpha})}\right)^{-1}
\left[\ln\frac{v_{f}}{A\left(L_{\beta}-L_{\alpha}\right)}+\frac{\epsilon_{0}}{k_{B}T}\right]
\end{equation}
Eq.~(\ref{SEq5}), which is Eq.~(11) of the Main Text, shows that the dependence of 
$f^{*}$ on $\ln v_f$ can be recast as $f^{*}$$=$$a\ln v_f$$+$$b$, with the slope
\begin{equation}
a=\left(\frac{z_{\alpha\beta}}{k_{B}T} - \frac{L_{\alpha}}{\kappa_{\alpha}
(L_{\beta}-L_{\alpha})}\right)^{-1}
\end{equation}
and the $y$-intercept
\begin{equation}
b = a\left[\frac{\epsilon_{0}}{k_{B}T}-\ln\left(A\left(L_{\beta}-L_{\alpha}\right)\right)\right]
\end{equation}

\section{Data analyses and modeling}

{\bf Constructing free-energy landscapes:} To build the energy landscape, we use the projection of the length of a single amino-acid residue as a reaction coordinate. For each model system, we collected the distribution $P(x)$ of the projections of residue lengths along the pulling axis $x$ (superhelical symmetry axis) and performed a plot the histogram using all structural snapshots saved. The landscape is then obtained by using the Boltzmann transformation, 
$G(x)$$=$$-k_BT$$\ln{P(x)}$. For constant force (force-clamp) MD simulations, we use 
the last 100~ns of trajectories to build the distributions. Using these simulations,
we devided each trajectory into three parts, i.e. when the system is in the $\alpha$-state, in the mixed $\alpha$$+$$\beta$-state, and in the $\beta$-state.

{\bf Estimating model parameters:} There are several parameters that
can be estimated either directly using the MD simulation output or using the energy landscape reconstruction described above. The length of the system in the $\alpha$-state $L_\alpha$ is an equilibrium distance between the two ends of the system. The spring constant $\kappa_\alpha$ is the tangent of the initial (elastic) part of the force-strain curve (in regime I) per residue length (unit length; see Fig.~1 and Fig.~S1). The maximum length in the $\beta$-state $L_\beta$ and the persistence length $l_\beta$ are computed from the worm-like chain fit to the non-linear part of the force-strain curve (regime III). To estimate the distance between the $\alpha$-state minimum and the peak of the energy barrier, $z_{\alpha\beta}$, and the distance between the peak of the barrier and the $\beta$-state minimum on the energy landscape, $z_{\beta\alpha}$, we first note that the positions of the minima and the barrier do not shift much when the force increases (see Fig.~3 and Fig.~S3). Hence, we can use the landscape constructed for $f>0$ by selecting the landscape for the system, e.g. in the mixed $\alpha$$+$$\beta$-state. The positions of the minima and the energy barrier are clearly visible on these landscapes, which allows for the estimation of the values for $z_{\alpha\beta}$ and $z_{\beta\alpha}$  (Fig.~3 and Fig.~S3). 

The energy landscape parameters can be estimated using the following approach. The Maxwell force, $f_0$, is the force at which the depths of the $\alpha$-state and $\beta$-state minima are equal. Out of $13$ constant force values used for myosin, $f$$=$$90~pN$ gives the closest values of energies at the minima, which are different by less than $0.1k_BT$ (see Fig.~S4 and {\it the inset} to Fig.~3). Hence, we take $f_0$$\approx$$90~pN$ for myosin. Similarly estimated, the value of $f_0$ for fibrin coiled-coils comes to $f_0$$\approx$$160~pN$ (Figs.~S3 and S4). The values for $\epsilon_{\alpha\beta}$ are computed from $f_0$ using Eq.~(9) in the Main Text. 
To obtain the values of $\epsilon_0$, we also use the energy landscape profiles
obtained for the forces at which both the $\alpha$-state and $\beta$-state are populated. Next, We measure the height of the barrier and then use the first 
Eq.~(7) in the Main Text, which reads:
$\Delta G_{\alpha\beta}(f)$$=$$\epsilon_0$$-$$f z_{\alpha\beta}$ to compute $\epsilon_0$. Finally, the attempt frequency $A$ is estimated as a fitting 
parameter (see Eq.~(6) in the Main text).

\section{Architecture of the coiled coils}
\label{sec:cc-arch}

A typical turn per amino acid residue in the $\alpha$-helix is around $100\degree$. Seven residues correspond to a $700\degree$ rotation which is $20\degree$ less 
than the $720\degree$ rotation for two complete turns. Hence, one can construct a seven-residue repeat that will lag $20\degree$ behind for each two turns. Since the $\alpha$-helices are right-handed, the aforementioned under-turn would give a left-handed stretch of residues of the surface of a straight $\alpha$-helix. In a 
coiled coil, when each $\alpha$-helix is twisted, this stretch would form a 
straight line along which two (of more) $\alpha$-helices interact with each other, forming a left-handed supercoil \cite{HarburyScience98}.

The seven-residue (or heptad) repeat that is usually denoted as 
{\bf a}-{\bf b}-{\bf c}-{\bf d}-{\bf e}-{\bf f}-{\bf g} (Fig.~\ref{fig:s5}), 
is a core feature of a so-called PV (``Peptide Velcro'') hypothesis \cite{ArndtStructure02,HarburyScience98}. This hypothesis separates residues in the heptad by their positions using the following assumptions. The residues {\bf a} and {\bf d} in the repeat are the residues that form hydrophobic core of the coiled coil. For example, in the double-stranded coiled coil the residues {\bf a} and {\bf d} from one of the $\alpha$-helices form hydrophobic contacts with residues {\bf a$^\prime$} and {\bf d$^\prime$} from the heptad repeat {\bf a$^\prime$}-{\bf b$^\prime$}-{\bf c$^\prime$}-{\bf d$^\prime$}-{\bf e$^\prime$}-{\bf f$^\prime$}-{\bf g$^\prime$} of the other $\alpha$-helix. The residues {\bf e} and {\bf g} are usually polar with long side chains, and they can further stabilize the coiled coils via the electrostatic interactions. Residues {\bf b}, {\bf c} and {\bf f} are hydrophilic since they form the exterior of the coiled coil (Fig.~\ref{fig:s5}).

\begin{figure*}[t]
\begin{center}
\includegraphics[width=1.0\linewidth]{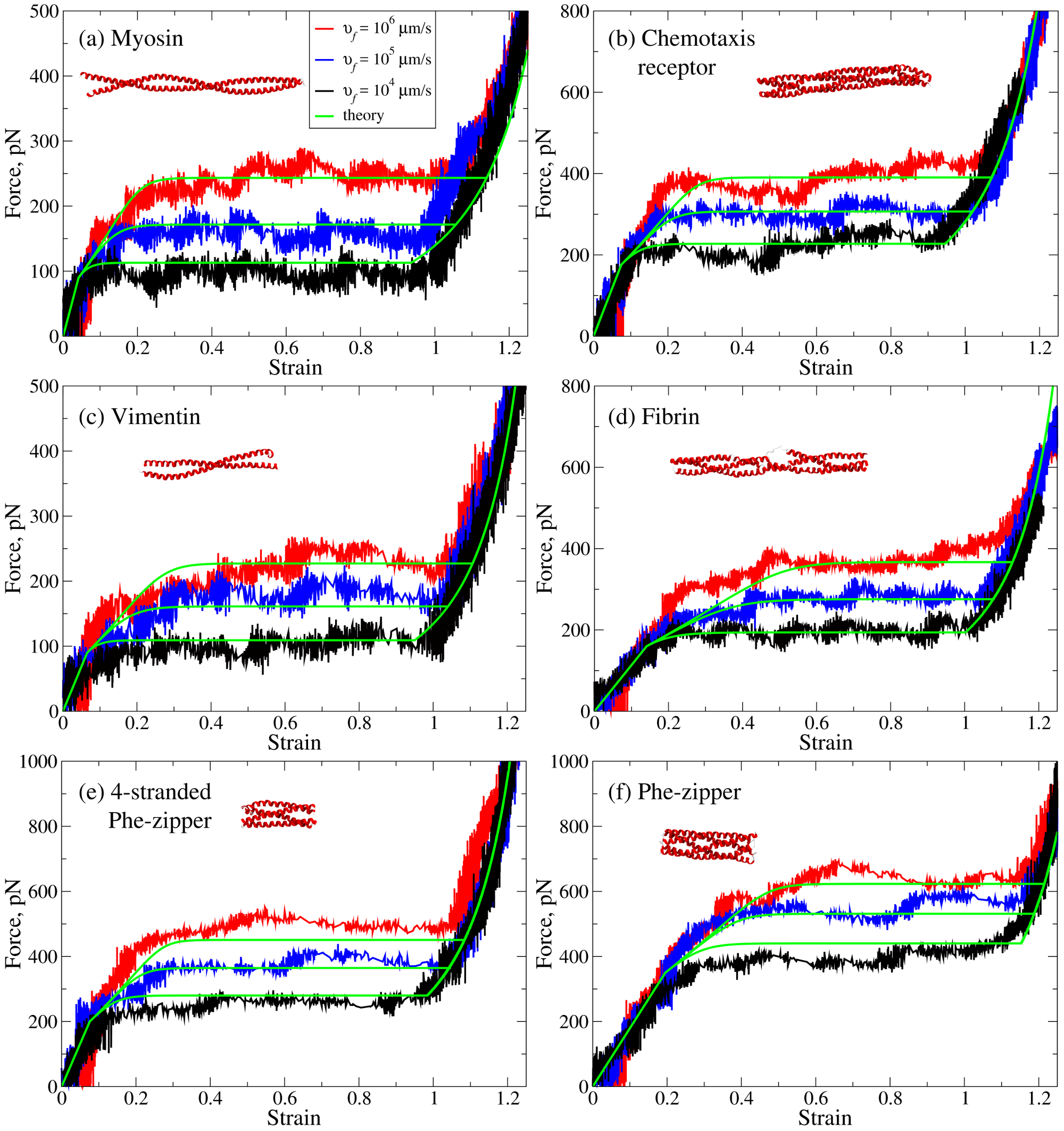}
\end{center}
\caption{Force-strain profiles ($f\varepsilon$-curves) for all superhelical protein fragments studied: myosin ({\it a}), bacterial chemotaxis receptor ({\it b}), vimantin ({\it c}), fibrin ({\it d}), four-stranded ({\it e}) and five-stranded ({\it f}) phenylalanine zippers. Detailed information about all the model systems is provided in Table~\ref{tab:systems}. The $f\varepsilon$-curves were obtained for various pulling speeds $v_f$$=$$10^4$, $10^5$ and $10^6$~$\mu$m/s (color 
code is presented in panel {\it a}). The dynamic force protocol is described in 
the Main Text. All the simulated $f\varepsilon$-curves show good agreement with the theoretical predictions obtained using continuum phase transition theory (green 
curves) described in the Main Text. All the model parameters are accumulated in 
Table~I in the Main Text).\label{fig:s1}}
\end{figure*}
\begin{figure*}[t]
\begin{center}
\includegraphics[width=1.0\linewidth]{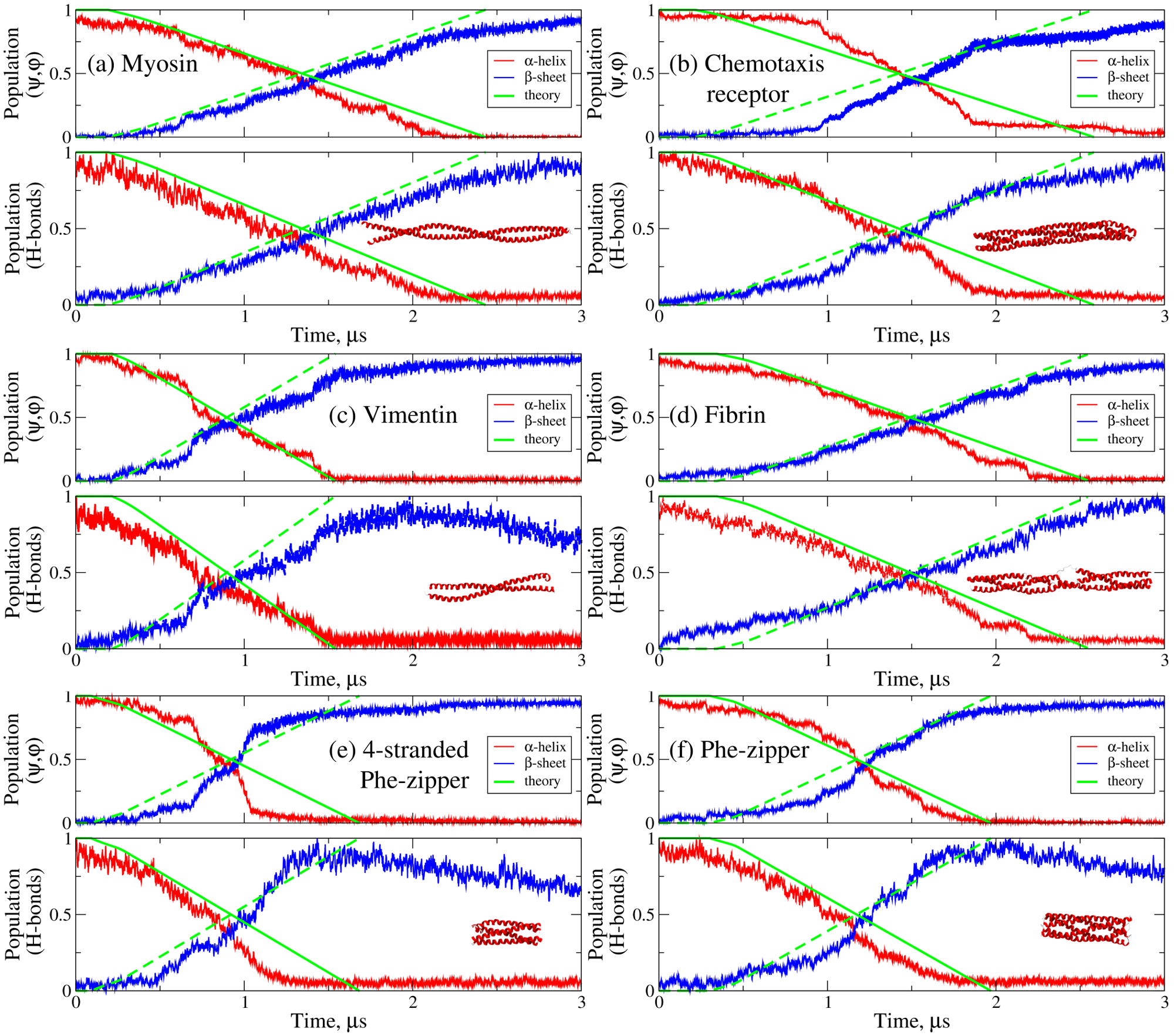}
\end{center}
\caption{Dynamics of the populations of the $\alpha$-state $p_{\alpha}$ and the $\beta$-state $p_{\beta}$ for all superhelical protein fragments studied: myosin
({\it a}), bacterial chemotaxis receptor ({\it b}), vimentin ({\it c}), fibrin ({\it d}),
four-stranded ({\it e}) and five-stranded ({\it f}) phenylalanine zippers. Detailed information about all the model systems is provided in Table~\ref{tab:systems}. The populations were obtained using the hydrogen bond 
(H-bond) based estimation (bottom panels) and the dihedral angles' based estimation 
(top panels) as described in the Main Text. (see also Fig.~2 in the Main text). Compared are the colorized curves from MD simulations (color code is explained 
in the graphs) and green (solid and dashed) curves obtained with continuum theory
described in the Main Text. For the H-bonds (D-H$\dots$A), we used the $3$$\AA$ 
cut-off for the distance between hydrogen donor atom D and acceptor atom A and the $20$$\degree$ cut-off for the D-H$\dots$A bond angle. For the dihedral angles, 
we used intervals $-80$$\degree$$<$$\phi$$<$$-48$$\degree$ and $-59$$\degree$$<$$\psi$$<$$-27$$\degree$ for the $\alpha$-phase, and $-150$$\degree$$<$$\phi$$<$$-90$$\degree$ and 
$90$$\degree$$<$$\psi$$<$$150$$\degree$ 
for the $\beta$-phase.\label{fig:s2}}
\end{figure*}
\begin{figure*}[t]
\begin{center}
\includegraphics[width=1.0\linewidth]{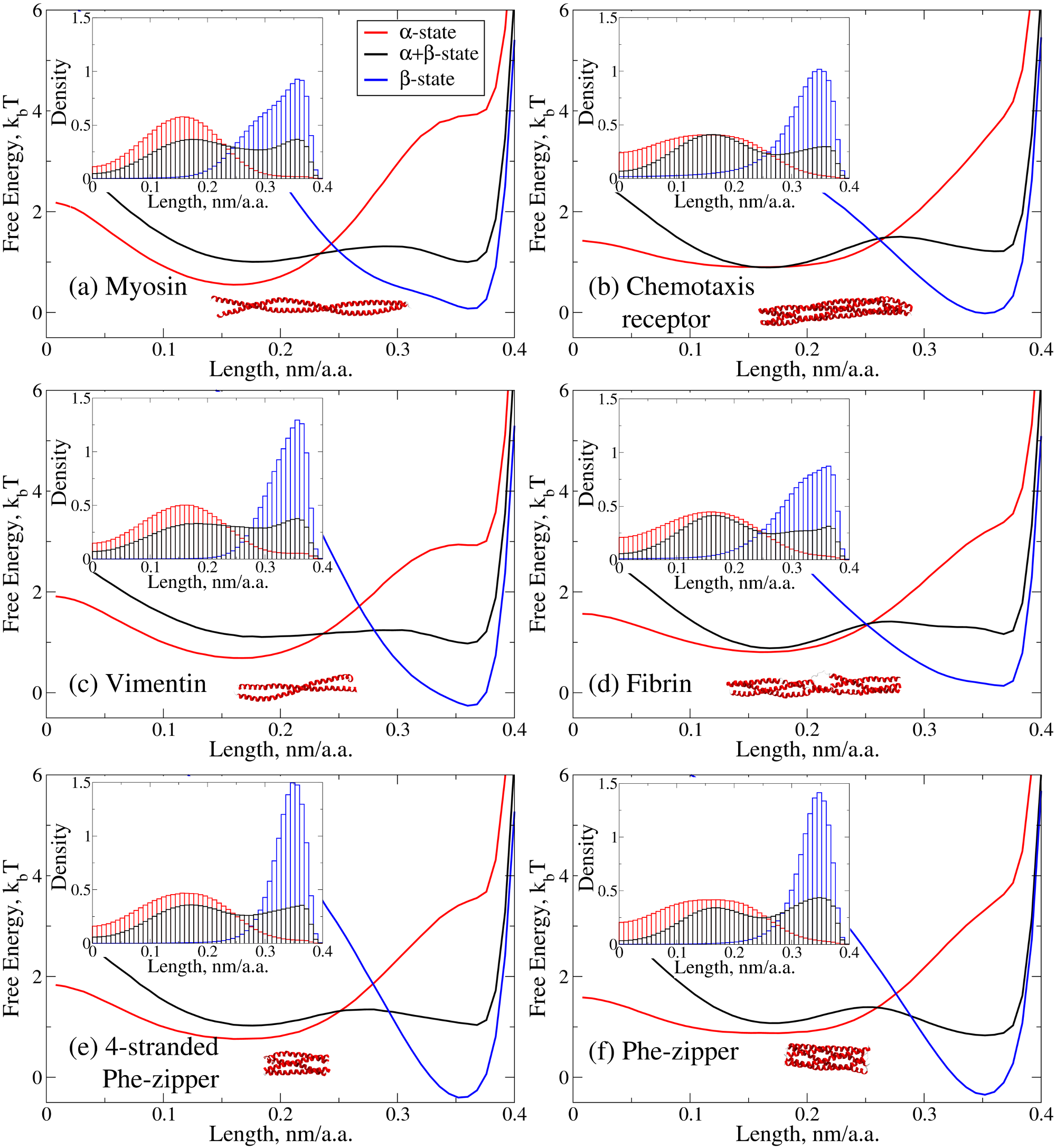}
\end{center}
\caption{Free-energy profiles as functions of the projection of amino acid residue length along the direction of pulling (energy landscapes) for all superhelical 
protein fragments studied: myosin
({\it a}), bacterial chemotaxis receptor ({\it b}), vimentin ({\it c}), fibrin ({\it d}),
four-stranded ({\it e}) and five-stranded ({\it f}) phenylalanine zippers. Detailed information about all the model systems is provided in Table~\ref{tab:systems}. For each model system, the landscapes are shown for the following values of constant pulling force: low force $f$$\ll$$f_0$ from the 
$0$$-$$60~pN$ (single left minimum corresponding to the $\alpha$-state; red curves); Maxwell force $f$$\approx$$f_0$ (two minima with the left (right) minimum corresponding to the $\alpha$-state ($\beta$-state); black curves; see Table~I for values of $f_0$ for all systems); and large force $f$$\gg$$f_0$ (single right minimum corresponding to the $\beta$-state; blue curves). {\it The insets} show the corresponding normalized histograms constructed based on the MD simulation output.\label{fig:s3}}
\end{figure*}
\begin{figure*}[t]
\begin{center}
\includegraphics[width=0.9\linewidth]{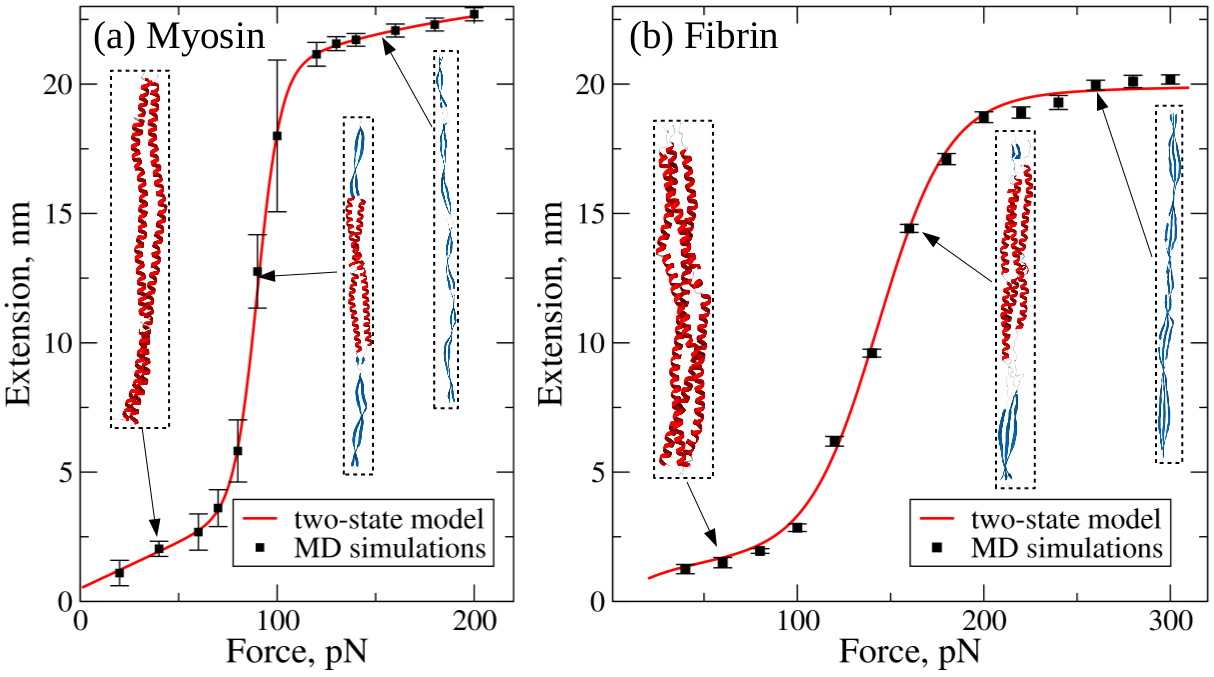}
\end{center}
\caption{The $\alpha$-to-$\beta$ structural transition in myosin superhelix ({\it a})
and fibrin superhelix ({\it b}) reflected in the sigmoidal profile of the average extension $\langle\Delta X\rangle$ (with standard deviations) versus constant 
pulling force $f$. The theoretical curves $X(f)$ (solid red line) was used to perform a fit to the elongation data from the dynamic force-clamp measurements {\it in silico} 
(black squares) as described in the Main Text. Shown are snapshots of the $\alpha$-phase, $\beta$-phase and mixed $\alpha$$+$$\beta$-phase. The phase diagrams were
constructed using the simulation output from five 200~ns long steered MD simulations (for each force value) for myosin and for fibrin. Theoretical curves were calculated
using the analytically tractable two-state model from our earlier study \cite{ZhmurovJACS12}.\label{fig:s4}}
\end{figure*}
\begin{figure*}[t]
\begin{center}
\includegraphics[width=0.9\linewidth]{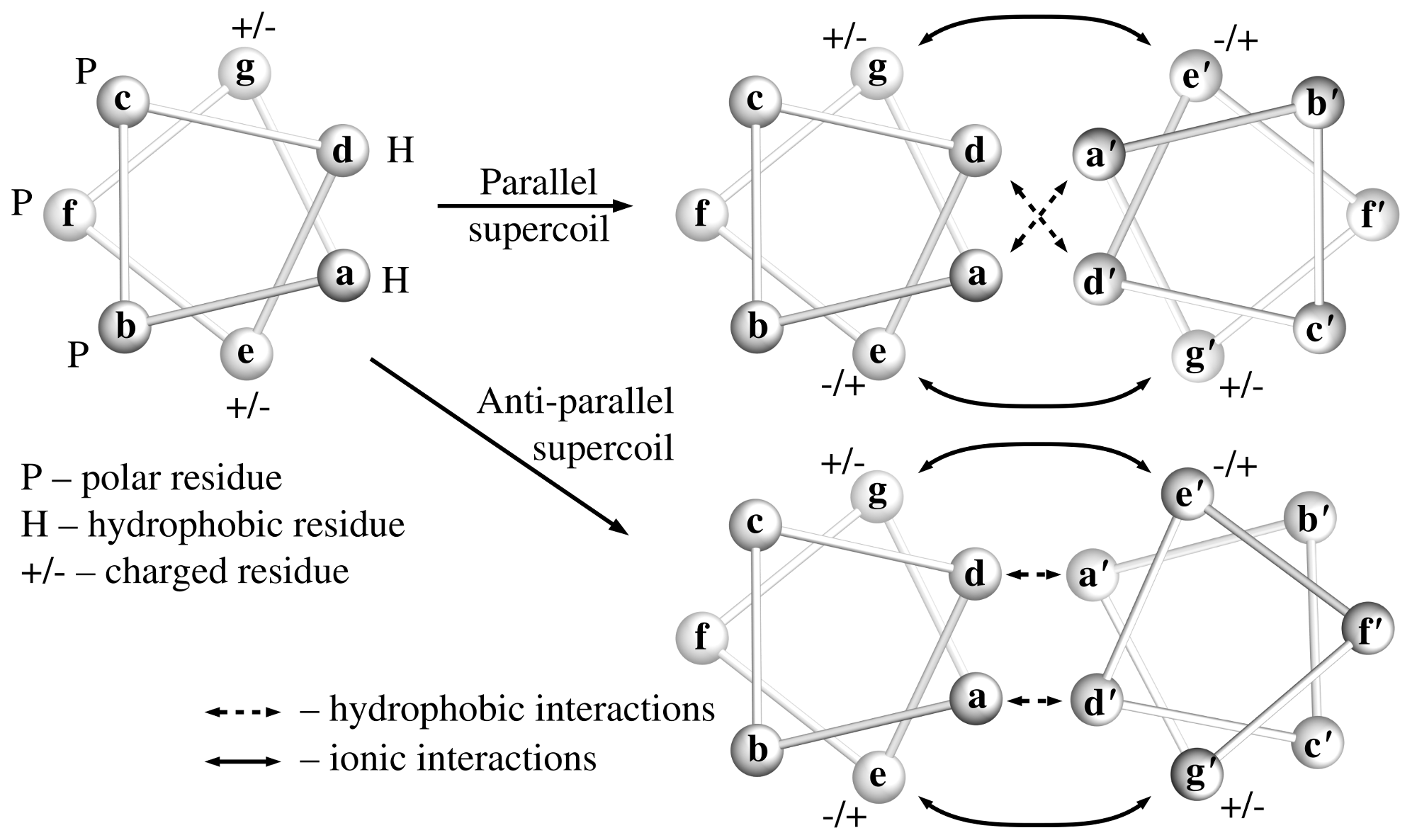}
\end{center}
\caption{Schematic representation of the $\alpha$-helices in superhelical proteins. In a heptad repeat of a left-handed coiled coil, residues {\bf a} and 
{\bf d} are hydrophobic. These residues form a hydrophobic core of 
the supercoil both in parallel and anti-parallel arrangements. The coiled coil is additionally stabilized by charged residues {\bf e} and {\bf g}. In a parallel supercoil, residue {\bf a} is close to residue {\bf a$^\prime$} and residue {\bf d} is close to {\bf d$^\prime$}. In an anti-parallel supercoil, residue {\bf a} forms a strong contact with residue {\bf d$^\prime$}, and residue {\bf d} forms a strong contact with residue {\bf a$^\prime$}. Supercoils with more than two $\alpha$-helices can be constructed in a similar fashion. For more information, see Refs.~\cite{HarburyScience98,MasonCBC04}.\label{fig:s5}}
\end{figure*}

\bibliography{coiledcoils}